\documentclass[preprint2]{aastex}

\usepackage{amsbsy}

\slugcomment{scheduled for 10 October 2001 Ap.J.}

\shorttitle{Theory of EGP Transits}
\shortauthors{Hubbard et al.}

\begin{document}


\title{Theory of Extrasolar Giant Planet Transits}


\author{W. B. Hubbard, J. J. Fortney, and J. I. Lunine}
\affil{Lunar and Planetary Laboratory, The University of Arizona,
    Tucson, AZ 85721-0092}

\email{hubbard, jfortney, jlunine@lpl.arizona.edu}

\and
\author{A. Burrows, D. Sudarsky, and P. Pinto}
\affil{Department of Astronomy and
 Steward Observatory, The University of Arizona,
    Tucson, AZ 85721}

\email{burrows, sudarsky, pinto@as.arizona.edu}



\begin{abstract}
We present a synthesis of physical effects influencing the observed
lightcurve of an extrasolar giant planet (EGP) transiting its host
star.  The synthesis includes a treatment of Rayleigh scattering,
cloud scattering,
refraction, and molecular absorption of starlight in the EGP
atmosphere.  Of these effects, molecular absorption dominates
in determining the transit-derived radius $R$, for planetary
orbital radii less than a few AU.  Using a generic
model for the atmosphere of EGP HD209458b, we perform a
fit to the best available transit lightcurve data, and infer
that this planet has a radius at a pressure of 1 bar, $R_1$,
equal to 94430 km, with an uncertainty of $\sim 500$ km
arising from plausible uncertainties in the atmospheric
temperature profile.  We predict that $R$ will be a function
of wavelength of observation, with a robust prediction of
variations of at least $\pm 1$\% at infrared wavelengths where H$_2$O
opacity in the high EGP atmosphere dominates.
\end{abstract}




\section{Introduction}

Measurements of the diminution of starlight during transit
of a planet across the
disk of a star provide an almost direct
means to detect
extrasolar giant planets (EGPs) with orbital
inclinations close to $90 \arcdeg$.
When coupled with measurements of the radial
velocity variation of the orbited star during
motion about the common barycenter,
the mass $M$ of the planet can be measured, and the
radius $R$ of the planet can be
deduced from the depth of the transit lightcurve.

To date, only one transiting extrasolar
planet has been observed: HD209458b
\citep{cha00,hen00}.  A
high-quality composite transit lightcurve has been obtained using the STIS
spectrograph on the Hubble Space Telescope (HST; Brown et al. 2000),
and a
model fit to the lightcurve and radial velocity data
yields the following results: inclination $i=86.68 \pm 0.14 \arcdeg$,
mass $M$=0.69 $M_{\rm J}$ (where $M_{\rm J}$ is the mass of Jupiter),
and radius $R=1.347 \pm 0.060 R_{\rm J}$
(where $R_{\rm J}$ is the radius of Jupiter).
Thus, HD209458b has been confirmed as a
genuine hydrogen-rich EGP \citep{bur00}.
Spectroscopic radial velocity data for HD209458 give
a precise value for the planet's orbital
period, $P$=3.524738 days.

\citet{bro00} modeled planet HD209458b as a uniform occulting disk of radius
$R$.
However, as \citet{ss00} first pointed out, the value of $R$
for a
real planet will be a function of wavelength, depending on the transmissive
properties of the planet's atmosphere, as well as on other properties of the
atmosphere, such as the
location of dense cloud layers.
In this paper, an unsubscripted $R$ will denote such a
wavelength-dependent radius, while subscripted $R$s will
denote values of the radius at a fixed level in the planet's
atmosphere.  Specifically, for purposes
of comparing the inferred values of $R$
and $M$ with theoretical models of EGPs of
specified age $t$,
one should relate the inferred $R$ to the radius of the planet at a
specific fiducial pressure, as is done for Jupiter, where $R_{\rm J}$
is customarily
expressed as $R_{\rm J}=71492 \pm 4$ km,
the equatorial radius at a pressure of 1 bar
\citep{lin81}. The purpose of the present paper,
then, is to fit the HST
lightcurve of \citet{bro00} to explicit atmospheric models of
HD209458b, in order to derive the EGP's radius at 1 bar pressure, which we will
call $R_1$.
Along the way, we obtain further predictions of the variation of $R$
with
wavelength, over a broader range of wavelengths than in the analysis of
\citet{ss00}.
Our model for the atmospheric structure is a generic one, but it
differs in some respects from that of \citet{ss00}.

The transit lightcurve depends on
(a) Rayleigh scattering of light from the host
star, (b) refraction of the stellar surface brightness distribution,
and (c) the slant
optical depth $\tau$ through the planet's atmosphere,
as determined by molecular
opacity and clouds.  All of these effects depend in turn upon the atmospheric
pressure ($P$) vs. temperature ($T$) profile, and upon the surface gravity $g$.

In the following, we consider the $P$ vs. $T$ profile, effects (a) through (c),
and
then present results for the relation of $R$ as a function of wavelength
$\lambda$ and the best-fit results for $R_1$ for HD209458b.

\section{Atmospheric Model}

\subsection{$P-T$ Profile}

Our philosophy in constructing the $P-T$ profile
of our baseline model
for HD209458b
is that this profile should be
representative for the planet's
albedo class (either Class IV or V),
as defined in \citet{sbp00}.
The specifics of this profile are not  
important as long as the basic molecular composition of the
atmosphere is respected and the mapping between pressure and areal
mass density is correct for a given gravity.  The surface gravity of 
HD209458b is measured to be close to the Earth's value. Hence, we used
the [T$_{\rm eff}$=1270 K/gravity=10$^3$ cm s$^{-2}$] model for Class IV
EGPs, similar to that found in SBP.  (Note that the actual
$P-T$ profile will be a function
of dynamic processes that redistribute heat from pole to equator and to  
the night side, processes that are difficult to model
[Guillot and Showman 2001].)  

Figure 1 shows the nominal $P - T$ profile used in the model (solid
heavy line
on left side), along with high-temperature and low-temperature versions
used to explore the sensitivity of the results to the profile.  At
$P$ = 1 mbar, the difference between the high-temperature and low-temperature
profiles is about 500 K, or about 50 percent of the nominal temperature.
As we discuss below, the transit radius of
HD209458b is primarily determined by
opacity sources in the vicinity of $P$ $\sim 10$ mbar.
In \S7 we discuss the sensitivity of the transit
radius to the alternative temperature profiles.

The right-hand side of Fig. 1 shows adiabatic $P - T$ profiles
for various interior models of the planet, to be discussed in
\S8.

\subsection{Clouds and Condensates}

We predict the altitude at which
clouds of various condensable species form using a code 
summarized in \citet{bur01} and \citet{mar99}.
Vapor pressure relations 
for the rocky condensates are from \citet{lun89}.
Above the altitude at which a 
given condensate first appears,
growth rates for particles and droplets are calculated using 
analytic expressions \citep{ros78}.
The cloud model assumes that the atmospheric 
thermal balance at each level is dominated by
a modal particle size that is the maximum 
attainable when growth rates are exceeded by the
sedimentation, or rainout rate, of the 
particles. In convective regions, equating the upwelling
(convective) velocity and the 
sedimentation velocity sets the particle size.  The
amount of condensate at each altitude 
within the cloud is given by the vapor pressure at
that level multiplied by a 
``supersaturation factor'' usually set to 0.01
by analogy with terrestrial clouds. The 
particle size and the mass density of the condensate
at each level thus determine the 
number density of particles.

The enstatite cloud layer is potentially the most important
for affecting the value of $R$.
Scattering and extinction cross sections for this major
cloud forming species
were obtained for the computed particle sizes by a
full Mie theory (SBP).
The optical properties of enstatite
were taken from \citet{dor95}.

\citet{aam01} develop a model of cloud formation that extends the 
foregoing to include a more realistic rainout prescription. However,
as shown in Fig. 2, 
the major cloud forming species for HD209458b, enstatite, does
not contribute opacity in 
the right altitude range to have a significant effect
on the transit profiles. More refractory 
cloud-forming species, e.g., aluminum silicates, occur
deeper in the atmosphere. Less 
refractory cloud formers such as water and sulfur-bearing
species would condense out at 
higher altitude, but the atmosphere is too warm for
these species to occur as clouds. 
Therefore, for HD209458b, our cloud model is more than
adequate. It is possible that 
very minor cloud forming species that are slightly less
refractory than enstatite would put 
some cloud opacity at modestly higher altitudes, but
their smaller abundance relative to 
enstatite would proportionately diminish their effect.
We therefore argue that, for this 
particular extrasolar planet and those with similar
effective temperatures, cloud opacity is 
not significant in determining the transit radius. 
Adopting the \citet{aam01} prescription would likely result in an even 
less significant role for the enstatite clouds, since such a model
would result in rainout of 
more condensate and lead to a less optically-thick cloud.

%% In this section, we use  the \subsection command to set off
%% a subsection.  \footnote is used to insert a footnote to the text.


%% This section also includes several of the displayed math environments
%% mentioned in the Author Guide.

\section{Rayleigh Scattering}

Rayleigh scattering was treated in an approximate manner.
As we demonstrate below, Rayleigh scattering has only a minor
effect on the value of $R$, so an approximate treatment is
sufficient.  The effect of Rayleigh scattering (or any other
scattering) is complex, because stellar photons incident on
the planet's atmosphere in a given pencil with incident
intensity $I$ are partially removed from the pencil
and scattered into different solid angles.  
Thus, if only conservative Rayleigh scattering were
occurring, the limb at $R$ would be
defined by the radius at which a high probability of such removal
occurs.  However, photons are scattered into the beam to the
observer as well as removed. Rather than treat the full
three-dimensional problem of Rayleigh scattering in a
spherically-stratified atmosphere, we replaced the atmosphere
with a series of slabs located in a plane containing the
center of the planet and orthogonal to the star-observer
line. In the following, we will denote the two-dimensional
vector separation of a point in this plane from the
projected planetary center by {\bf r}, and the scalar
value by $r$.  The three-dimensional vector position
of an atmospheric point from the planetary center
will be denoted by {\bf r$''$}.

Each slab, located at a two-dimensional radius $r$ from the projected
planet center, has a Rayleigh-scattering optical depth
$\tau_{\rm R}$ given by
\begin{equation}
\tau_{\rm R}=2 \int_r^{\infty} r'' dr''
\sigma_{\rm R} N(r'') / \sqrt{r''^2 - r^2},
\end{equation}
where $N$ is the number density of molecules in the atmosphere,
and $\sigma_{\rm R}$ is the Rayleigh-scattering cross-section
per molecule, given by
\begin{equation}
\sigma_{\rm R}={{8 \pi^3 (2+\nu)^2 \nu^2} \over {3 \lambda^4 N^2}},
\end{equation}
where $\nu$ is the refractivity (refractive index minus 1) of the gas
\citep{cha60}.
Since $\nu \ll 1$ and $\nu \propto N$, $\sigma_{\rm R}$ is
a function only of $\lambda$ and the gas composition.  

We wrote a Monte Carlo scattering
code to investigate the effects of Rayleigh scattering in the
planet's atmosphere.  The code
follows every photon as it travels through a plane-parallel
slab of a given optical thickness.  Incident photons can arrive
at the top of the slab from any direction on an imaginary hemisphere.
Inside the slab, after every scattering event, a new photon direction in
three dimensions is calculated, based on the Rayleigh-scattering phase
function.  Photons are followed until they emerge from either side of
the slab, and the total path traveled, in units of optical depth, is
tabulated.  When the photon finally emerges from the bottom of the slab
it is placed into a bin that corresponds to its final direction.  Upon
completion, the total number of photons in each bin is obtained, along
with the average path traveled per photon in that bin.

Our code was tested against the analytical results of \citet{vdh74}
and \citet{cha60}.
To within the noise in
the Monte Carlo simulations ($\sim 1$\%),
we were able to match van de Hulst's results for an isotropic
distribution of photons and Chandrasekhar's for isotropic and ``pencil
beam'' distributions at a variety of incident angles.

We ran our simulations for radiation at normal incidence to the slab for
a variety of different optical thicknesses, logarithmically spaced from
0.01 to 28.  If we assume an imaginary observer looks at normal incidence
from the other side of the slab, the observer will see photons that pass
through the slab unobstructed,
as well as those that are multiply scattered
back into the beam and emerge normal to the surface.  Of most interest
to us in this situation are those photons that, although scattered,
emerge normal to the slab surface and ultimately reach the
observer.  These photons create a Rayleigh-scattering
``glow'' from the slab.  We found that the glow
intensity in the direction of the observer
increases until $\tau_{\rm R} \sim 3.4$, as fewer
and fewer photons are able to pass
through the slab unscattered.  At $\tau_{\rm R} > 3.4$,
up to $\tau_{\rm R}=28$, the greatest
thickness we ran, Rayleigh-scattered intensity decreases more or less
exponentially, as a larger fraction of the photons 
are scattered back out
the top of the slab, rather than scattering all the way through.  We fitted
an equation to both sides of this curve so that the
glow intensity could be
interpolated, and extrapolated to higher $\tau_{\rm R}$
if necessary.  We also fitted
an equation to the average optical path length traveled for these
photons, using this result to estimate the total molecular-absorption
optical depth, $\tau_{\rm M}$, and cloud optical depth,
$\tau_{\rm C}$.  The total optical depth for photons
initially incident with impact parameter {\bf r} is then given by
$\tau=\tau_{\rm R}+\tau_{\rm M}$+$\tau_{\rm C}$.
In practice, for all cases that
we have investigated, $\tau_{\rm M}$ becomes large long before
$\tau_{\rm R}$ or $\tau_{\rm C}$ does.  Likewise,
refractive effects are still negligible when
$\tau_{\rm M}$ becomes significant, as discussed below.
Thus, for purposes of computing $\tau_{\rm M}$, the photon path
can be taken to be a straight line through the planetary
atmosphere, such that
\begin{equation}
\tau_{\rm M}=2 \int_r^{\infty} r'' dr''
\sigma_{\rm M} N(r'') / \sqrt{r''^2 - r^2},
\end{equation} \label{taum}
where $\sigma_{\rm M}$ is the average absorption cross-section per
molecule for a solar-composition planetary atmosphere.
We estimate $\tau_{\rm C}$ with a similar formula, although
strictly speaking,
it becomes appreciable at levels where refraction is important.

Since in our transit calculations the planet is a disk passing in front
of its star, we assume the atmosphere of the planet is a flat slab with
an optical thickness that decreases with increasing distance from the
planet center.  In this way, our results for different slab thicknesses
can be used directly in our planetary atmosphere calculations.

In our transit simulations, the atmosphere of the planet was broken up
into many small annuli, each with its own optical thickness, which is
dependent on wavelength, as calculated using our model for the structure
of the planet's atmosphere.  At a given wavelength of light, for every
optical thickness in the atmosphere, the Rayleigh glow intensity was
calculated.  This intensity results in a small additive component
to the star plus planet signal outside of transit,
but in general has no detectable effect on the
observed light curve.  In addition, since the total average path
traveled for a scattered photon is also known, the effects of absorption
due to molecular opacity can be calculated.  At high optical
thicknesses, most scattered photons that would emerge from the slab are
actually absorbed by opacity sources in the atmosphere.

%% This section contains more display math examples, including unnumbered
%% equations (displaymath environment). The last paragraph includes some
%% examples of in-line math featuring a couple of the AASTeX symbol macros.

\section{Refraction}

%% The displaymath environment will produce the same sort of equation as
%% the equation environment, except that the equation will not be numbered
%% by LaTeX.

The theory used to compute refractive effects is essentially
identical to that of the standard theory for occultations of
stars by planetary atmospheres \citep{hyl90}.
In the following, we let $I({\bf r})$ denote
the photon intensity within a differential solid angle whose coordinates
are labeled by the two-dimensional vector {\bf r} measured in
the plane of the sky, from the center of the planet.
The intensity $I({\bf r})$ of the observed image of star plus planet
is then given by the mapping
\begin{equation}
I({\bf r}) = I'({\bf r'}) e^{-\tau(r)},
\end{equation}
where $I'({\bf r'})$ is the stellar surface brightness
distribution, $\tau$ is the total optical depth integrated along a
ray path with impact parameter $r$, and ${\bf r'}$ and {\bf r}
are two-dimensional vectors in a plane normal to the
propagation direction, marking the starting point of a ray on
the stellar surface and its closest-approach position in the
planet's atmosphere, respectively.

The mapping from ${\bf r'}$ to {\bf r} is obtained by
computing the total phase shift $\Phi$ imposed by the refractivity
distribution on a photon with impact parameter $r$,
\begin{equation}
\Phi(r)=(4 \pi / \lambda) \int_r^{\infty} r'' dr''
\nu(r'') / \sqrt{r''^2 - r^2},
\end{equation}
then computing the two-dimensional bending angle ${\boldsymbol \alpha}$
according to ${\boldsymbol \alpha} = (\lambda / 2 \pi) \nabla \Phi$,
where the gradient is taken in the two-dimensional plane.

Figure 2 shows the calculated values of $\tau_{\rm M}$,
$\tau_{\rm R}$, and cloud optical depth, $\tau_{\rm C}$,
for our best-fit planetary model, as discussed in \S6.
Figure 2 also shows the atmospheric radius where
refraction begins to be important.  Specifically, the shaded
region in Fig. 2 shows the difference between $r$ and $r'$
at a level where the difference,
\begin{equation}
r-r'=D\alpha,
\end{equation}
becomes equal
to 500 km, or about one atmospheric scale height ($D$ is
the distance from the star to the planet).  The upper
edge of the shaded region corresponds to $r$ and the lower
edge to $r'$.  Since $\alpha$ varies
exponentially with $-r$, considerable ray bending occurs
for impact parameters below the shaded region.  However,
$\tau_{\rm M}$ is always so large in this region that
refraction is unimportant for defining $R$, for
values of $D$ similar to that of HD209458b.

We do not include general-relativistic ray bending in the
theory presented in this paper, as it is unimportant for
the parameters of HD209458b.  However, it is straightforward
to include gravitational lensing.  One simply adds to the
refractive $\alpha$
discussed above, an additional term for general-relativistic
ray bending,
$\alpha_{\rm GR} = 4GM/rc^2$ (where $G$ is the gravitational
constant, $M$ is the planet's mass, and $c$ is the speed of
light).  The total bending angle used in Eq. (6) is then
$\alpha + \alpha_{\rm GR}$.

\section{Gaseous Opacities}

The primary gaseous absorptive opacity sources in the atmospheres of
hot EGPs include H$_2$, H$_2$O, CH$_4$, CO, and the
important alkali metals, Na and K.  We take the temperature-
and pressure-dependent opacities from 
theoretical and experimental data referenced in 
\citet{bur97} and \citet{bur01}.
In the near infrared ($\sim$1 to 2.6 $\mu$m), absorption by
H$_2$O molecules figures most prominently, with strong ro-vibrational
bands centered at $\sim$0.95, 1.15, 1.4, 1.85, and 2.6 $\mu$m. The
strong pressure-broadened resonance lines of neutral Na and K -- the
strengths of which depend on the level of ionization by stellar
irradiation -- appear prominently at $\sim$0.59 and 0.77 $\mu$m,
respectively.  The dominant carbon-bearing molecule is a function of both
temperature and pressure.  At lower temperatures, CH$_4$ is dominant,
but CO overtakes CH$_4$ at higher temperatures ($\sim$950 K at 0.1 bar,
$\sim$1100 K at 1 bar).  Hence, the strengths of CH$_4$ features (1.4,
1.7, 2.2 $\mu$m) and CO features (1.2, 1.6, 2.3 $\mu$m) are highly
temperature-dependent.  Finally, an important continuous opacity source
in cloud-free EGPs is H$_2$-H$_2$ collision induced absorption at
high pressures and temperatures \citep{zhe95}.
  
All of the opacity sources mentioned above were used to
calculate $\sigma_{\rm M}$, and then incorporated in
Eq. (\ref{taum}); the results are plotted in Fig. 2.

\section{Transit Lightcurve}

The next step in calculating a transit lightcurve was to
synthesize images of the two-dimensional distribution of
starlight around the planet.  We created a synthetic square
aperture of size $361 \times 361$ pixels (1 pixel = spatial
scale of 700 km), centered on the planet.  The star was
taken to be a disk with a pixel intensity of $1$ at its center,
with a prescribed darkening law to the limb.
At each pixel in the array, the pixel intensity was calculated
using the calculated values of the various $\tau$s, adding
the Rayleigh-scattered component into the beam and the
refracted stellar component, from each pixel on the stellar
disk (including contributions from virtual pixels on the
part of the stellar profile outside the synthetic aperture).
The total pixel sum over the aperture was then calculated,
with and without the planet present, allowing us to calculate
the total intensity subtracted and added by the planet.

We then synthesized a transit lightcurve, incorporating all the physical
effects described above, and using most of the fitted parameters of
\citet{bro00}: $i$, stellar radius, and orbital radius and period. 
The lightcurve was obtained by averaging over the bandpass of
the \citet{bro00} experiment, weighted by a blackbody distribution
for the effective temperature of HD209458.  The value of
$R_1$ for HD209458b was adjusted in our model until we
matched the depth of the theoretical transit lightcurve
to the depth of the composite observed lightcurve, as
shown in Fig. 3.
The match of our model, which has only the adjustable
parameter $R_1$, to the data is quite good.
We obtain $R_1=94430$ km for HD209458b, for the nominal
$P-T$ profile shown in Fig. 1.  For the cold profile,
we obtain $R_1=95000$ km, and for the hot profile,
$R_1=94260$ km.

The flux from the star was computed using limb-darkening coefficients
from \citet{vnh93}.
The two equations used were the nonlinear
logarithmic and square-root laws.  These laws are
(logarithmic):
\begin{equation}
I(\mu)/I(1) = 1 - A (1 - \mu) - B  \mu  \ln(\mu),
\end{equation}
and (square root):
\begin{equation}
I(\mu)/I(1) = 1 - C  (1 - \mu) - D  (1 - \sqrt{\mu}),
\end{equation}
where $I(1)$ is the specific intensity at the center of the stellar disk,
$A$, $B$, $C$, and $D$ are wavelength-dependent constants, and
$\mu$ is the cosine of
the angle between the line of sight and the emergent intensity.

The curve shown in Fig. 3 is for the logarithmic law; we found
that use of the square root law instead
changed the lightcurve by less than 1 part in
400.  The coefficients used were interpolated from the Van Hamme
monochromatic coefficient tables, for $\log g = 4.3$, $T_{\rm eff}$
of 6000 K, and
solar metallicity.  The very slight mismatch with the data is perhaps
due to a difference in the calculated stellar intensities, which are
based on the stellar models of \citet{kur91},
and the actual change in
stellar intensity across the disk.

Figures 4 and 5 show examples of synthetic images and
provide more details about the
contributions of Rayleigh scattering and refraction
to the shape of a transit lightcurve,
which are generally negligible
for the specific case of HD209458b.  In Fig. 4a,
we show that even at a short wavelength of 0.45 $\mu$m,
the forward-scattered Rayleigh glow is almost invisible.
In Fig. 4b, we reduce $\tau_{\rm M}$ by a factor $10^3$
and stretch the image to make the narrow Rayleigh
glow annulus more apparent.

In Fig. 5, we retain the atmospheric structure for
the nominal model corresponding to Fig. 4(a), but
we increase the distance $D$ to
$10^4$ times the value for HD209458b, thus bringing the
refracting layers of the atmosphere into play.
According to Eq. (6), for $r - r^{\prime}$ fixed at,
say, one scale height ($\sim 500$ km), then
if $D$ increases by $10^4$, $\alpha$ must decrease
by the same factor.  Since $\alpha$ is roughly proportional
to pressure in the atmospheres considered here, a decrease
in $\alpha$ by this factor corresponds to displacing
the layer responsible for the prescribed refraction upward by about
nine scale heights, to a region where the slant optical
depth $\tau$ is negligible.  The atmospheric depth
where the limb of bright refracted starlight terminates
is determined by the abrupt increase of $\tau$ to appreciable
values.

As $D$ increases further,
our theory smoothly transforms to conventional ray-optical stellar
occultation theory in the limit $\tau=0$
\citep{hyl90}, appropriate to observation
of the passage of a planet in front of a star of smaller apparent
angular size than the planet. 

Except for such conventional stellar occultations,
we are unlikely to observe
a transit dominated by refractive effects, as the probability of
a transit by an EGP at an orbital radius of hundreds of AU
is exceedingly small. 

\section{Variation of $R$ with $\lambda$}

As discussed above, the relation $R(\lambda)$ is almost entirely
determined by molecular absorption features.  \citet{ss00}
discussed a possible dramatic variation of $R(\lambda)$ due to
alkali absorption features in the visual wavelength bands.
We predict dramatic effects at infrared wavelengths as well,
due to strong features of H$_2$O; however, detection
of infrared variations in $R(\lambda)$ may require observations
above the Earth's atmosphere.

Figure 6 shows, for HD209458b parameters,
the $R(\lambda)$ relation predicted by our three
$P-T$ models.  This relation is computed by evaluating binned
averages of opacities (for specified temperatures)
at 500 individual frequencies spaced
across several $\times 10^4$ original frequencies.  Note that
the variation of $R$ vs. $\lambda$ increases with $T$ at some
wavelengths, and decreases at others.  Note also
that though a flattened temperature gradient in the upper
atmosphere will smooth out the reflection/emission
spectrum of the planet itself, the variation
$R(\lambda)$ with wavelength will not be similarly
flattened.  For example, even though the Na/K alkali
metal features in the planet's spectrum may be smoothed by
stellar irradiation, the transit size will still vary appreciably
across these features as long as Na/K are not ionized.

Because the variations with wavelength are quite rapid in some
wavelength intervals, it may be possible to strategically
choose wavelengths of observations which will span these
rapid variations and still be close enough together to allow
the limb-darkening of HD209458 to be removed by fitting a smooth
model.  Note that $R$ is typically about 2000 km larger than
$R_1$, and thus pertains to atmospheric layers which are at
pressures near $\sim 10$ mbar.

\section{Conclusions}

Referring back to Fig. 1, note that an adiabat in a solar-composition
object with a mass of 0.69 $M_{\rm J}$, which would yield a value
of $R_1$ compatible with the one determined here, must have a
significantly lower specific entropy than the atmospheric layers
heated by the star.  It follows that there must be a significant
region in the planet, possibly spanning pressures from
$\sim 10$ to $\sim 10^4$ bar,
where the $T$ vs. $P$ relation must be substantially subadiabatic
or even isothermal.  Detailed thermal evolution models for this
layer remain to be calculated.

We have shown that apparent variations of at least $\pm 1$\%
in the radius of giant planets occur 
as a function of wavelength. This variation is potentially discernable with the next 
generation of transit-observing spacecraft in Earth orbit, provided
that they possess the 
capability of multi-wavelength observation with
sufficient spectral resolution. We 
therefore recommend that proposers or designers of such missions consider the 
possibility of transit-based probing of extra-solar planet atmospheric composition, via 
multi-wavelength observations. 

Finally, as is suggested by Fig. 6, there could be considerable
variation in $R(\lambda)$ at ultraviolet wavelengths.
Indeed, as is well known \citep{smh90}, the location
of the level where $\tau \sim 1$ in our own giant planets
is a strong function of ultraviolet wavelength,
and in the UV
is typically at much lower pressures than 1 $\mu$bar.
Observation of an EGP transit at UV wavelengths
is, in many experimental aspects,
equivalent to a solar-occultation UV
experiment in our solar system, and may have similar
diagnostic power for chemical composition of
the high planetary atmosphere.  To be sure,
Jupiter has a rather warm stratosphere/mesosphere,
with photochemical aerosols, so the UV radius could be a sensitive function
of the external boundary conditions within which 
an EGP planet finds itself.

\acknowledgments

We would like to thank Tim Brown and David Charbonneau
for an advanced look at the HST/STIS data, and Lam Hui and
Sara Seager for a copy of their draft paper on refractive effects
on transits, which we received while this paper was in
review.
This research was supported by Grants
NAG5-7211, NAG5-7499, and NAG5-10629
(NASA Origins Program), NAG5-4214 (NASA Planetary Astronomy),
and NAG5-7073 (NASA Astrophysics Theory).

\clearpage


\begin{figure}
\plotone{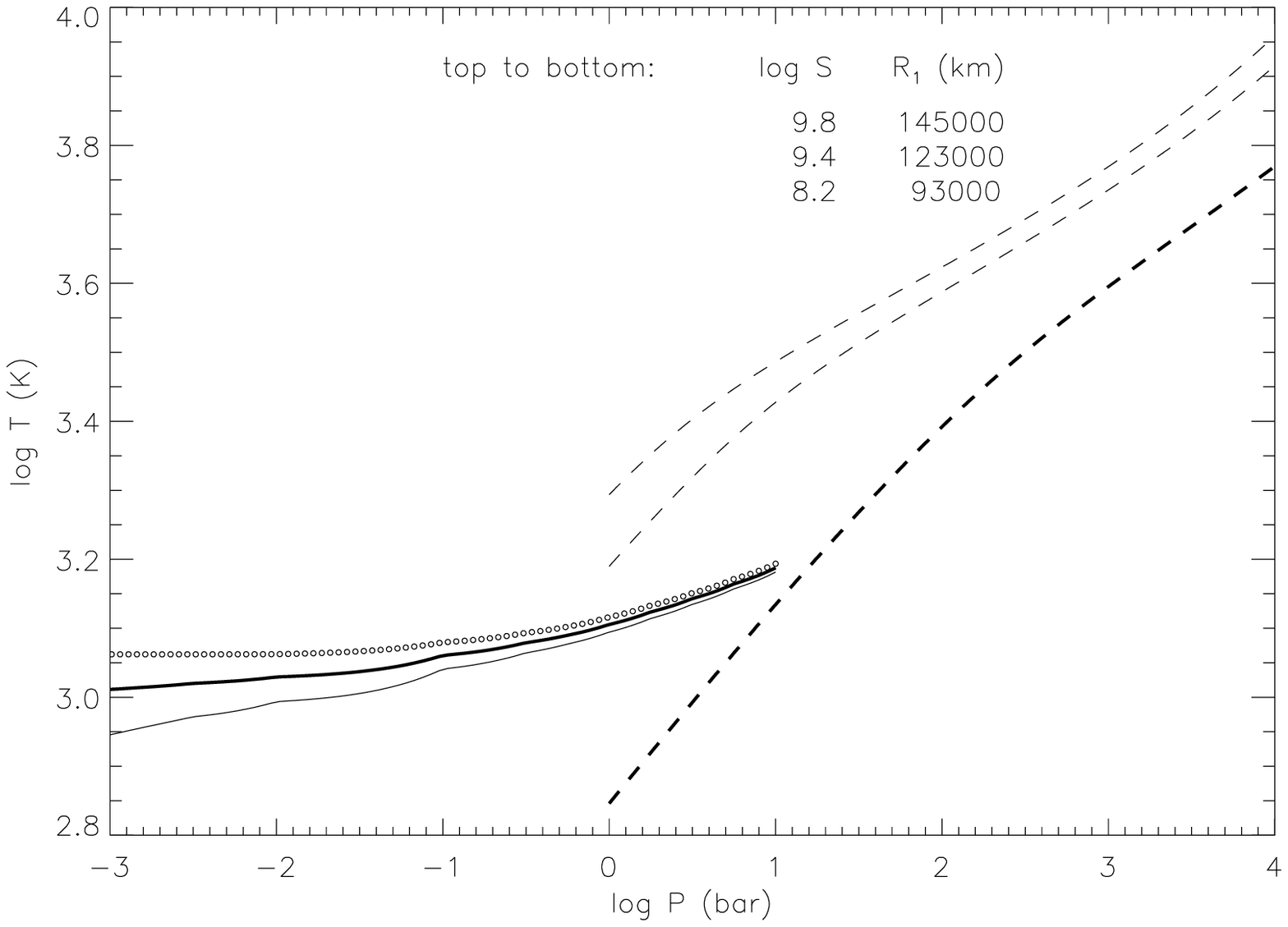}
\caption{{\it Heavy line} on the left shows the nominal atmospheric
pressure-temperature profile used in these calculations;
{\it light solid line} shows the low-temperature version and
{\it open-dotted line} shows the high-temperature version.
{\it Dashed lines} show interior H-He adiabats, with the corresponding
entropy (in units of Boltzmann's constant per baryon) and
total radius $R_1$ for a planet mass of 0.69 $M_{\rm J}$.
{\it Heavy dashed line} shows the approximate interior adiabat that
matches the transit-derived value of $R_1$. \label{fig1}}
\end{figure}

\clearpage 

\begin{figure}
\plotone{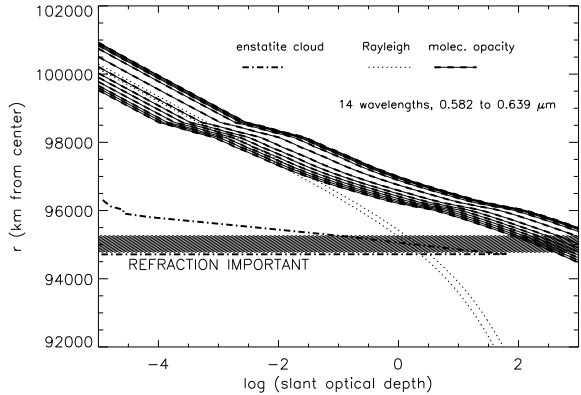}
\caption{Radius vs. optical depth for $\tau_{\rm M}$ ({\it dashed}),
$\tau_{\rm R}$ ({\it dotted}), and $\tau_{\rm C}$ ({\it dot-dashed}), for
the nominal $P-T$ profile and for parameters of HD209458b.
Below the {\it shaded zone} near 1 bar pressure, refraction is
also significant. Molecular opacity profiles are plotted
for wavelengths spanned by the observations of Brown et al.
(2000). \label{fig2}}
\end{figure}

\begin{figure}
\plotone{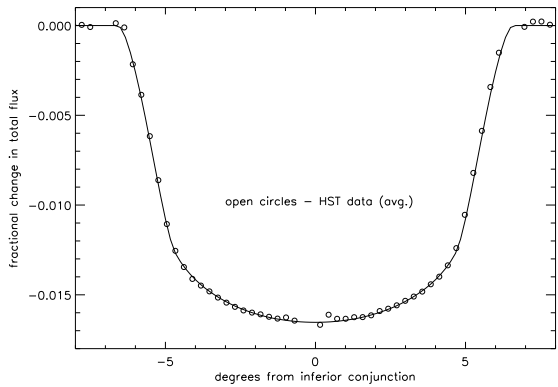}
\caption{Synthetic transit lightcurve for our model
({\it solid curve}), compared with composite lightcurve of Brown et al. (2000).
The three $P-T$ profiles give indistinguishable solid curves.
 \label{fig3}}
\end{figure}

\begin{figure}
\plottwo{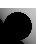}{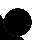}
\caption{(a) Synthetic image of a transit of HD209458b
(pixel scale is from 0 to 1, with 1 corresponding to the
intensity at the center of the stellar disk), with the planet
at an orbital position $5 \arcdeg$
from inferior conjunction, at a wavelength of
0.45 $\mu$m. Close scrutiny will show a faint Rayleigh-scattering
ring around the planet's limb exterior to the stellar limb.
(b) An image of the same geometry, but with molecular
opacity reduced by a factor 1000, and with the pixel scale
stretched over the range 0 to 0.1. The Rayleigh-scattering
ring is thus wider and brighter.  \label{fig4ab}}
\end{figure}

\begin{figure}
\plotone{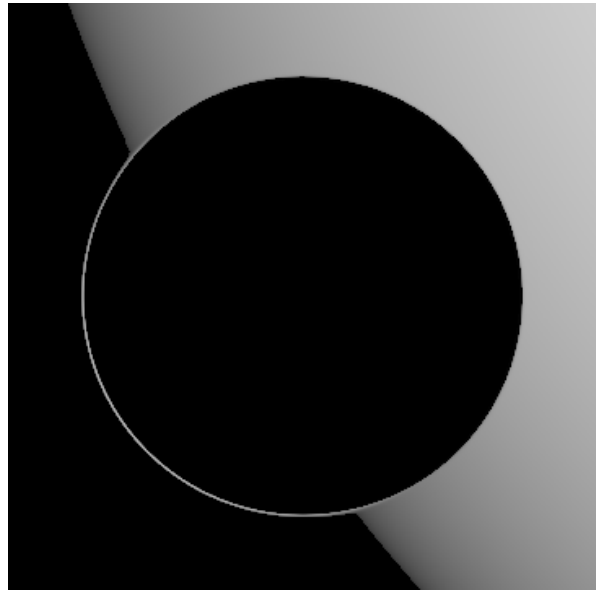}
\caption{Image of the same atmosphere as in Fig. 4(a)
(pixel scale is again from 0 to 1), but with the distance $D$
from the planet to the star increased from the
value for HD209458b to a value $10^4$ times greater.
The bright limb outlining the planet's disk is far brighter
than the Rayleigh ``glow'', and is produced by refractive
mapping of the stellar disk's intensity distribution. \label{fig5}}
\end{figure}

\begin{figure}
\plotone{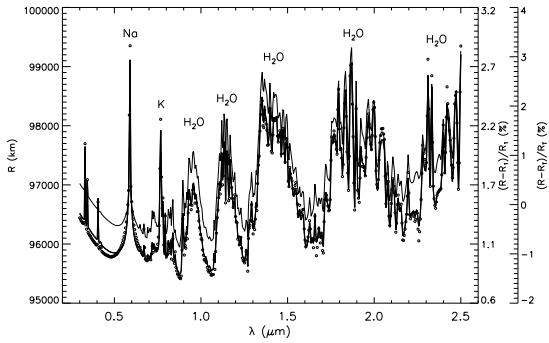}
\caption{Predicted variation of transit radius $R$
with wavelength ({\it heavy curve}, nominal $P-T$ profile;
{\it light curve}, cold $P-T$ profile; {\it open dots},
hot $P-T$ profile; see Fig. 1). The right-hand scales show, in percent,
the variation of $R$ with respect to $R_1$ and with
respect to $R_{\rm T}$, an $\sim$``average'' transit radius in
the visual wavelength band, adopted as
$R_{\rm T}=96500$ km $=1.35 R_{\rm J}$.  At wavelengths
where slant optical depth is high, $R$ is larger.
``Absorption'' features thus appear upside-down on
this plot.  Prominent features are labeled with the
responsible molecule.  \label{fig6}}
\end{figure}

%% If you are not including electonic art with your submission, you may
%% mark up your captions using the \figcaption command. See the 
%% No more than seven \figcaption commands are allowed per page, 
%% so if you have more than seven captions, insert a \clearpage 

%% The following command ends your manuscript. LaTeX will ignore any text

\end{document}